\documentclass[12pt,preprint]{aastex}

\renewcommand {\deg}   {\mbox{$^\circ$}}

\newcommand   {\kms}   {\mbox{km\,s$^{-1}$}}
\renewcommand {\ga}    {\mbox{\rlap{\hbox{\lower5pt\hbox{$\sim$}}}\hbox{$>$}}}
\renewcommand {\la}    {\mbox{\rlap{\hbox{\lower5pt\hbox{$\sim$}}}\hbox{$<$}}}

\begin{document}
\pagenumbering{arabic} 
\def\kms {\hbox{km{\hskip0.1em}s$^{-1}$}} 
\voffset=-0.8in

\def\msol{\hbox{$\hbox{M}_\odot$}}
\def\lsol{\hbox{$\hbox{L}_\odot$}}
\def\kms{km s$^{-1}$}
\def\Blos{B$_{\rm los}$}
\def\etal   {{\it et al.}}                     
\def\psec           {$.\negthinspace^{s}$}
\def\pasec          {$.\negthinspace^{\prime\prime}$}
\def\pdeg           {$.\kern-.25em ^{^\circ}$}
\def\degree{\ifmmode{^\circ} \else{$^\circ$}\fi}
\def\ut #1 #2 { \, \textrm{#1}^{#2}} 
\def\u #1 { \, \textrm{#1}}          
\def\nH {n_\mathrm{H}}
\def\ddeg   {\hbox{$.\!\!^\circ$}}              
\def\deg    {$^{\circ}$}                        
\def\le     {$\leq$}                            
\def\sec    {$^{\rm s}$}                        
\def\msol   {\hbox{$M_\odot$}}                  
\def\i      {\hbox{\it I}}                      
\def\v      {\hbox{\it V}}                      
\def\dasec  {\hbox{$.\!\!^{\prime\prime}$}}     
\def\asec   {$^{\prime\prime}$}                 
\def\dasec  {\hbox{$.\!\!^{\prime\prime}$}}     
\def\dsec   {\hbox{$.\!\!^{\rm s}$}}            
\def\min    {$^{\rm m}$}                        
\def\hour   {$^{\rm h}$}                        
\def\amin   {$^{\prime}$}                       
\def\lsol{\, \hbox{$\hbox{L}_\odot$}}
\def\sec    {$^{\rm s}$}                        
\def\etal   {{\it et al.}}                     
\def\la{\lower.4ex\hbox{$\;\buildrel <\over{\scriptstyle\sim}\;$}}
\def\ga{\lower.4ex\hbox{$\;\buildrel >\over{\scriptstyle\sim}\;$}}
\def\refitem{\par\noindent\hangindent\parindent}
\oddsidemargin = 0pt \topmargin = 0pt \hoffset = 0mm \voffset = -17mm
\textwidth = 160mm  \textheight = 244mm
\parindent 0pt
\parskip 5pt

\shorttitle{Sgr B2}
\shortauthors{}

\title{An OH(1720 MHz) Maser and a  Nonthermal  Radio Source in\\
Sgr B2(M): A  SNR--Molecular Cloud Interaction Site?}

\author{F. Yusef-Zadeh$^1$, W. Cotton$^2$, M. Wardle$^3$, \& H.  Intema$^4$}
\affil{$^1$CIERA, Department of Physics and Astronomy, Northwestern University, \\Evanston, IL 60208}
\affil{$^2$National Radio Astronomy Observatory,  Charlottesville, VA 22903}
\affil{$^3$Department of Physics and Astronomy,  Macquarie University, Sydney NSW 2109, Australia}
\affil{$^4$Leiden Observatory, Leiden University, P.O. Box 9513, NL-2300 RA Leiden, The Netherlands}

\begin{abstract} 
Sgr B2 is a well-known star forming molecular cloud complex in the Galactic center region
showing   evidence of high energy activity as traced by 
the K$\alpha$ neutral FeI line at 6.4 keV,  
as well as GeV and TeV $\gamma$-ray emission.  
Here we present VLA and GMRT  observations with respective  resolutions of 
$\approx3.5''\times1.2''$ and 25$''\times25''$  
and report  the detection  of an  OH(1720 MHz) maser, 
with no accompanying OH 1665, 1667 and 1612  MHz maser emission.  
The maser coincides  with a 150 MHz nonthermal radio source in Sgr B2(M). 
This  rare class of OH(1720 MHz) masers or the so-called supernova remnant (SNR)  masers, 
with no main line transitions,  
trace shocked gas  and signal the interaction of an expanding SNR with a molecular cloud. 
We interpret the 150 MHz radio source as either the site of a
SNR -- molecular gas interaction  or a wind-wind collision in  a massive binary system. 
The interaction of the molecular cloud and  the nonthermal source enhances 
the cosmic-ray ionization rate, allows   
the diffusion of cosmic rays into  the cloud and 
produces  the variable 6.4 keV line, GeV and TeV $\gamma$-ray emission from Sgr B2(M). 
The cosmic ray electron  interaction
with the gas in the Galactic center 
can not only explain  the measured  high values of cosmic ray
ionization and heating rates but also 
contribute to nonthermal
bremsstrahlung continuum emission,  all of which 
are consistent with observations. 
\end{abstract} 



\section{Introduction}

Interaction between a supernova remnant and a molecular cloud is traced by 
bright spots of 1720 MHz OH  
maser emission (Frail, Goss \& Slysh 1994; Wardle \&  Yusef-Zadeh 2002). About 24 known supernova remnants 
 are associated with this rare class of masers 
(Frail \etal\, 
1996; Green \etal\,  1997; Yusef-Zadeh \etal\, 1996, 2003a,b; Green \etal\,  1997; 
Koralesky \etal\,  1998; Hewitt \& Yusef-Zadeh  2009; Brogan \etal\, 2013). These OH(1720 MHz) masers are
collisionally pumped in molecular gas at temperatures and density range  
50-200 K and $\sim10^{4-6}$\,  cm$^{-3}$, respectively (Lockett, 
Gauthier \& Elitzur 1999). 
The enhanced column of OH at the interaction site 
requires X-rays or  cosmic rays from the SNR to  
irradiate the water-rich gas behind a C-type shock wave 
(Wardle, Yusef-Zadeh \& Geballe 1999; 
Wardle 1999).  SNRs interacting with molecular clouds have also shown a correlation between 
$\gamma$-ray and X-ray emission as well as  SNR masers (Yusef-Zadeh \etal\, 2003a; Hewitt, Yusef-Zadeh and Wardle 2009), 
thus consistent with an   enhanced cosmic ray flux or X-rays. 


One site that might be suitable to study molecular shocks and high energy emission in the Galaxy is Sgr B2, 
a spectacular star forming region in the 
Galactic center. Sgr B2 is associated with a massive molecular cloud 
and contains 41 UC  HII regions; 
(e.g., Lis \& Goldsmith 1989; 
De Pree \etal\, 2014).
Recent monitoring of UC HII regions indicate significant radio continuum flux variation over 
the last 20 years (De Pree \etal\, 2014). 
Sgr B2 is also a prominent site of high energy activity, being a  
a source of hard X-rays, GeV and TeV $\gamma$-rays. It also displays  variable fluorescent 
emission in the 
neutral iron 
K$\alpha$ line  at 6.4 keV (Revnivtstev \etal\, 2004; Terrier \etal\, 2010) 
either due to irradiation by a  burst of X-rays  from Sgr A* (e.g., Zhang \etal\, 2015) or  
 a variable  low energy cosmic-ray electron flux (Yusef-Zadeh \etal\, 2007a). 
In fact, a radio source was identified toward Sgr B2 at 255 and 327 MHz, 
suggesting that it is a 
nonthermal  source embedded in the cloud and 
that the origin of the FeI K$\alpha$ line emission is tied to 
a   high cosmic ray ionization rate in the cloud (Yusef-Zadeh \etal\, 2007b).   

  
To search for sites of  interaction between SNRs and molecular gas, we carried out a sensitivity--limited OH 
survey of the region between Sgr C (G0.5-0.0) and Sgr B2 (G0.67--0.05) of the Galactic center.  
We find  only two clouds from  which  OH(1720 MHz) 
masers are detected. One is associated with the  
Sgr A East SNR (G0.0+0.0) interacting with the 50 \kms\, molecular cloud (e.g., 
Wardle, Yusef-Zadeh \& Geballe 1999; 
Yusef-Zadeh \etal\,   1996; 
Sjouwerman \& Philstr\"om 2008). The other is the detection of  an   OH(1720 MHz) maser 
coincident with Sgr B2(M)  with no  accompanying 
1665/67 MHz emission. Here we focus on the  relationship between the OH(1720 MHz) maser 
and a  150 MHz nonthermal source found in Sgr B2(M).  

\section{Observation and Data Reduction: VLA \& GMRT}

A survey of the Galactic center was carried out
with  the Jansky Very Large Array (VLA)
searching for all four ground state transitions of OH (1612, 1665,  1667 and  1720 MHz). 
Details of these observations,  
which covered  the inner $96'\times 43'$ ($l \times b$) using 11 pointings along the inner Galactic plane,  
will be given elsewhere (Cotton \& Yusef-Zadeh 2016, in preparation). 
Briefly, the L-band  observations were centered on the rest frequency of the
OH satellite  (1.612231, 1.72053 GHz)
and main (1.66540184, 1.667359 GHz) lines  and were made on May 31, 2014 using the VLA in ``A''
configuration. 
The velocity range covered was $\approx -420$ to +235 \kms\, with 
a spectral and spatial resolution of $\approx$ 0.73 \kms\, and  
$\approx3.5''\times$1.2'', respectively. 
Each pointing was observed for 2 minutes in three scans separated by
about 45 minutes.
Calibration followed the procedure given in Cotton \& Yusef-Zadeh (2016).


The 1.7 GHz data were imaged using faceting to cover the primary
beam.  Continuum imaging was done after phase self-calibration on each pointing using all four subbands
out to a radius of 30$'$  and was CLEANed. 
In the quieter fields, the off source RMS in the continuum images was
$\approx$1 mJy beam$^{-1}$ although with much variation and many artifacts from
the strong, almost completely resolved emission. 
Spectral line channels were imaged to a radius of 13.6$'$. 
Typical RMSes in line free channels are $\approx$50 mJy beam$^{-1}$.


Archival data from a 150 MHz GMRT sky survey (TGSS) 
were  obtained from the GMRT archive, and processed with a fully automated pipeline 
(Intema \etal\, 2009, 2014), which includes direction-dependent calibration, modeling and imaging of 
ionospheric phase delay. 
A full survey of the radio sky at 150 MHz as visible from the GMRT (Swarup \etal\, 1991) was performed 
between 2010 and early 2012, covering the full 
declination range of $-55$ to $+90$ degrees. 
The survey consists of more than 5000 
pointings on an approximate hexagonal grid. Data were  recorded in half polarization (RR,LL) every 2 seconds, in 256 
frequency channels across 16 MHz of bandwidth (140--156 MHz). 
Each pointing was observed for about 15 minutes, 
split over 3 or more scans spaced out in time to improve UV-coverage. 




\section{Results}

Apart from the known OH(1720 MHz) masers associated with the Sgr A East SNR (Yusef-Zadeh \etal\, 1996; 
Sjouwerman \& Philstr\"om 2008), the surveyed region uncovers  only one  additional OH(1720 MHz) source, 
G0.6647-0.0358,  
in Sgr B2(M).  Figure 1 shows the spectrum  of OH(1720 MHz) emission toward 
this source in Sgr B2. A list of 
fitted parameters of OH(1720 MHz) emission  is given in Table 1. Entries in columns 1 to 9 give the source 
name,  RA, Dec coordinates, Galactic l, b coordinates, peak flux density, radial velocity, the linewidth and 
the lower limit to the brightness temperature value, respectively. 
The velocity  of the  OH(1720 MHz) maser is $\sim62$ \kms\, corresponding to the systematic velocity of the 
cloud. 
G0.6647-0.0358  is a maser 
because of its  narrow linewidths $\sim0.7$ \kms\, and a lower limit to the brightness temperature 
T$_b\sim4.8\times10^4$ K  
assuming a  source size of $3.5''\times1.2''$.
We also searched for OH emission at 1665, 1667 and 1720 MHz from the position of 
G0.6647-0.0358 and 
found  a 
1$\sigma$ RMS noise of $\sim200$, 100 and 35
mJy beam$^{-1}$ at 1665, 1667 and   1612 MHz, respectively  (Cotton and 
Yusef-Zadeh 2016).


A number of past studies have detected strong OH maser emission 
from G0.6647-0.0358 in Sgr B2(M) (e.g., 
Caswell \& Haynes 1983; Gaume \& Mutel 
1987; Argon, Reid \& Menten 2000 (ARM)).  
A sensitive OH survey of the sky with the VLA 
with a spectral resolution of 0.14 \kms, 
spatial resolution of 2.6$''\times1.5''$ and RMS sensitivity of 90 mJy channel$^{-1}$
detected  two  OH (1720 MHz)  masers 
 within   1$\sigma$ positional  error of G0.6647-0.0358 (ARM).
Both  masers, which are detected in two studies, 
 are highly polarized and have flux densities ranging  between 1.6 and 14 Jy with 
linewidths between  0.38 and 0.46 \kms\, (ARM; Gaume \& Mutel 1987)
Given our poorer sensitivity and spectral resolution, 
it is most likely that the  two masers are spectrally not resolved in our low-resolution  data and/or are  variable.
ARM 
found no 1612/1665/1667 MHz counterparts to the  OH(1720 MHz) masers in Sgr B2(M) 
whereas 
Gaume \& Mutel (1987)  find a OH(1665 MHz) LCP counterpart with a flux density of 0.4 Jy.  
The OH(1720 MHz) maser G0.6647-0.0358 is displaced from star forming OH masers in Sgr B2(M) by  about $3''$ 
and is likely to be    a member  of a rare class 
of OH masers related to SNR masers. However, 
there is a compact HII region Sgr B2 D that is identified at 1.3cm 
within 0.5$''$ of  the OH(1720 MHz) maser  (De Pree et al.  1996), thus we can not rule out the possibility 
that  the  OH(1720 MHz) masers are associated with Sgr B2 D.


\begin{deluxetable} {rrrrrrrrrr}
\label{OH1720MaserCatalog}
\tablecaption{Properties of the fit to an OH(1720 MHz) Maser in Sgr B2}
\tablewidth{0pt}
\tablehead{
\colhead{\small Name} & \colhead{\small RA (J2000)} & \colhead{\small Dec (J2000)} &\colhead{\small G. long} & \colhead{\small G. lat} & \colhead{\small Flux Density} & 
\colhead{\small Vel}& \colhead{\small Width} & \colhead{\small T$_b$}\\
               & \colhead{\small $17^{\rm h}\, 47^{\rm m}$} &  \colhead{\small $-28^\circ\, 23'$}   & \colhead{\small $\circ$} & \colhead{\small $\circ$}   & \colhead{\small mJy} 
&  \colhead{\small km s$^{-1}$} & \colhead{\small km s$^{-1}$} & \colhead{\small $^\circ$K} 
}
\startdata
\scriptsize{G0.6647-0.0358} & \scriptsize{20$^s.03\pm0.10$} & \scriptsize{$12''.38\pm0.11$}  & \scriptsize{0.6647} &  \scriptsize{--0.0358} &  \scriptsize{749$\pm64$}  & \scriptsize{61.3} & \scriptsize{0.7} & \scriptsize{4.8E04}
 \\
\enddata
\end{deluxetable}

Figure 2a  shows contours of 150 MHz emission  with a resolution of 25$''\times25''$ 
superimposed on a grayscale image at 23 GHz with  a resolution of 
$\sim0.3''\times0.2''$ (PA$\sim68^\circ$) (De Pree \etal\, 2005).  
The 150 MHz  source also coincides with a 
255 and 327 MHz source within  resolutions of $22.5''\times16.8''$  and 
$12.6''\times6.8''$, respectively (Yusef-Zadeh, Wardle \& Roy 2007). 
Figure 2b shows contours of 327 MHz from Sgr B2(M) based on VLA observations (Nord \etal\, 2004). 
The cross drawn on these figures  coincides with  the position of the  OH(1720 MHz) maser G0.6647-0.0358
at the  velocity of  $\sim$ 62  \kms.  
The concentration of compact and UC HII regions traced at 23 GHz 
lie to the NE of the 327 and 150 MHz peak emission. 
This is because the emission from  these HII regions become  opaque at low frequencies.   
The offset between 327 and 150 MHz peak emission, noted by comparing contours of emission in Figure 2a and 2b, 
is explained by the contribution of 
thermal emission to the nonthermal source  at 327 MHz. 
The color image in Figure 3a shows the composite image at 150 MHz, 327 MHz and 1.4 GHz. 
The extended  feature (blue) traces the emission from  HII regions at high frequency. 
A close-up of the region where low frequency emission dominates,  is shown in Figure 3b. 
The emission at  255 and 327 MHz 
shows   weak extended  emission  from thermal HII regions whereas there is no evidence of 
weak HII emission at 150 MHz. 

We estimate the contribution of HII regions to the flux density at six different frequencies 
within a 25$''$ beam at the position of the remnant,  
$S_\nu = (S_\mathrm{SNR}+S^{(1)}_\textsc{h\,ii}+S^{(2)}_\textsc{{\footnotesize hii}}) \exp(-\tau_s)$ with emitted
flux densities $S_\mathrm{SNR}$ from the remnant, and $S^{(1)}_\textsc{h\,ii}$, $S^{(2)}_\textsc{h\,ii}$ from compact and
UC  HII regions within the beam, respectively.  Here $\tau_s= \tau_{150}\,(\nu/150\,\mathrm{MHz})^{-2.1}$ is
the optical depth of the extended ionized screen lying in front of the Galactic center, which has $\tau_{150}\sim 1$
(Roy 2013).
 The remnant is assumed to have a non-thermal spectrum $S_\mathrm{SNR} = S_{150}
(\nu/150\,\mathrm{MHZ})^{-\alpha}$.  
Additional  support for nonthermal emission from Sgr B2 comes from GBT observations (Hollis \etal\, 2007; Crocker 
\etal\, 2007 but see Lang, Palmer \& Goss 2008).
The flux densities  from the HII regions are written $S^{(i)}_\textsc{h\,ii} = f
^{(i)}\,\Omega_B\,B_\nu(T)\,(1-\exp(-\tau^{(i)}_\nu))$ where $f^{(i)}$ is the beam filling factor, $\Omega_B$ is the
beam solid angle, $B_\nu(T)$ is the Planck function, and $\tau^{(i)}_\nu = (\nu/\nu^{(i)}_t)^{-2.1}$ is the optical
depth and $\nu_t$\, is the turnover frequency.

A formal fitting of this model to the data has not been made because there is correlation  amongst the parameters.  
Instead, we show an example in Figure 4, obtained by setting $\tau_{150}=1$, $\alpha=0.5$, $T=8500\,\mathrm{K}$ and
adjusting the remaining parameters to obtain a reasonable match to the observed spectrum, finding $S_{150}=0.30$\,Jy,
$ f^{(1)}=0.22$, $\nu^{(1)}_t=2.3$\,GHz, $ f^{(2)}=2.2\times10^{-3}$, and $\nu^{(2)}_t=50$\,GHz.  Note that the 1.4
and 4.8\,GHz flux densities  constrain the flux density and turnover frequency of the compact HII region component
 but the parameters
for the UC  HII region component are less constrained, as the beam filling factor and turnover frequency can
be traded off to yield the observed 23\,GHz flux density.

The spectral index of the SNR is poorly constrained because of the uncertain magnitude of the absorption by the 
foreground screen.  For example, a similar fit to the low-frequency data can be obtained adopting a flat SNR spectrum
($\alpha=0$) and setting $\tau_{150}\approx 0.5$.  Nevertheless, in this case $S_{150}\approx 0.18$\,Jy and we
conclude that the remnant's intrinsic flux density is $\sim 0.25$\,Jy at 150\,MHz even though its spectral index is poorly
constrained.

\section{Discussion}

The most interesting result of our observations is the evidence for OH(1720 MHz) masers coincident with a low 
frequency nonthermal radio source in Sgr B2(M). The detection of a 150 MHz source and an OH(1720 MHz) maser at the 
radial velocity of Sgr B2 provide   compelling evidence for shocked molecular gas resulting 
from the interaction of a nonthermal radio source and the Sgr 
B2(M) molecular cloud. This rare class of OH masers 
is different than star forming OH masers where 
all four transitions of OH are observed as  masers and the 1665/67 OH masers are strongest. 
A number of star formation OH masers are detected in the Sgr B2(M)  
but there is no OH(1720 MHz) detected  within 3$''$ of G0.6647-0.0358 (see Fig. 4 of ARM). 
The characteristics of the OH(1720 MHz) maser in Sgr B2  are similar to those observed in the Galaxy where SNRs 
expand into a molecular cloud, interact with the gas, increase the cosmic ray ionization rate at the site of the 
interaction.  
The nonthermal source with an OH(1702 MHz) counterpart is likely to be the site of an interaction of 
an expanding remnant with the Sgr B2 cloud. 
It is possible that remnant is extended beyond G0.6647-0.0358  
with  low  surface brightness, thus is not detectable with present 150 MHz sensitivity. 

Given the large number of UC HII region excited by young O stars in Sgr B2,  the progenitor to  
the nonthermal source is likely to be  an O star that exploded as a supernova, 
and we are witnessing a  shock from an expanding SNR  driving into the parent molecular cloud.  
Another possibility is 
that the nonthermal emission is produced  in a massive binary system 
(e.g. Chapman \etal\,  1999; Eichler \& Usov 1993), though to date 
all the  known OH(1720 MHz) masers are found in interacting SNRs.



\subsection{High Energy Emission from Sgr B2(M)}

The evidence for a compact SNR interacting with the molecular gas has  a number of important implications. First, 
population inversion and significant amplification in the 1720 MHz line requires densities $\sim 10^5$ cm$^{-3}$, temperatures in 
the range 30--125 K, an OH column density in the range $10^{16}$--$10^{17}\ut cm -2 $ and an absence of a strong FIR 
continuum (Lockett, Gauthier \& Elitzur 1999).  
In this picture, the interacting region must be self-shielded against 
FIR radiation; 
the UV radiation from young massive stars is absorbed by dust grains and reradiated in the 
FIR. 
Second, the synchrotron source injects relativistic particles 
into the molecular cloud at the interaction site. 
In particular, low energy cosmic rays must contribute   to the ionization and 
heating of molecular gas potentially  producing the 
FeI 6.4 keV line and hard X-ray emission, 
 whereas high energy cosmic ray particles interact with the dense gas,  yielding  
$\gamma$-ray emission.

We use the observed 150 MHz flux density to estimate the likely magnitude of these effects. Unfortunately the spectral index 
is poorly determined, so we consider E$^{-2}$  or E$^{-3}$ 
relativistic electron spectra, which would yield $\nu^{-0.5}$ or $\nu^{-1}$ 
synchrotron spectra, respectively. 
We assume that the electron spectrum extends between 1 MeV and 1 GeV, that the 
proton to electron ratio is 100 and  adopt a spherical source with radius 1 pc.
Then we find an equipartition magnetic field of 0.26 or 0.6 mG for  E$^{-2}$ or E$^{-3}$, respectively. 
This implies an energy 
density of relativistic electrons $\sim$ 17 or 88 eV cm$^{-3}$. 
The total energy in relativistic electrons is 
$\sim4.2\times10^{44}$\, or
$\sim2.2\times10^{45}$\, erg. 
This  implies that cosmic ray electrons ionization rate is 
$\zeta\sim1.1\times10^{-14}$\, or 
$\zeta\sim2.5\times10^{-13}$\, 
s$^{-1}$ (see equation 3 of Yusef-Zadeh \etal\, 2013).  
This estimate is consistent with a 
 number of recent measurements 
indicating  a vast amount of diffuse H$_3^+$,  OH$^+$, H$_2$O$^+$ and 
H$_3$O$^+$ distributed in Sgr B2 and 
the Galactic center,  implying 
a cosmic ray ionization rate $\zeta\sim 10^{-15}\,-10^{-14}\, \rm s^{-1}\, \rm H^{-1}$ in the Galactic center
region is  
one to two orders of magnitude higher 
than in the Galactic disk (Oka et al. 2005; Geballe \& Oka 2010; van der Tak \etal\,  2006; 
Indriolo \etal\, 2015; Le Petit \etal\, 2016). 

The Sgr B2 cloud is the most massive cloud of the Galactic center region  
with hydrogen   column density   N$_H\sim9.8\times10^{24}\,  \rm cm^{-2}$ toward Sgr B2(M) (Lis \& Goldsmith  1989; 
Etxaluze \etal\, 2013). 
The interaction of relativistic  electrons and their secondaries with the molecular gas also produces Fe K$\alpha$ line 
emission and a nonthermal bremsstrahlung continuum extending from X-rays to $\gamma$-rays. 
To estimate the K$\alpha$ line 
emission,  we note that each ionization is associated with the loss of 40eV of energy by relativistic electrons and 
that the efficiency of K$\alpha$  photon production is $\sim$200 photons  for each erg of energy lost 
(Yusef-Zadeh et al 2013).
Using the 
values of the cosmic ray ionization rate and the column density, the intensity of K$\alpha$ line emission
at 6.4 keV is estimated  
$I_{K\alpha}\sim1.3\times10^{-6}$\, 
or 
$\sim2.9\times10^{-5}$\, 
photons s$^{-1}$\, cm$^{-2}$\, arcmin$^{-2}$, assuming a metalicity of 3 (see equation 6 of Yusef-Zadeh \etal\, 2013). 

The  interaction of  cosmic ray electrons
with the gas in the Galactic center can not only explain  the measured  high values of cosmic ray
ionization and heating rates but also 
contribute to nonthermal
bremsstrahlung continuum emission. 
Assuming    the number density of hydrogen nuclei in atomic or
molecular form, 
n$_H$  = n(HI) + 2n(H2) $\sim3\times10^6$\, cm$^{-3}$, 
the expected $\gamma$-ray flux at 1 GeV and 1 TeV are  
$F_{\gamma} \approx 3\times10^{-10}\,$ and $3\times10^{-19}\,  {\rm \ ph \ cm^{-2} \ s^{-1} \ GeV^{-1} }$, 
respectively, 
for $\alpha=1$ or 
$\approx 1\times10^{-8}\,$ and $1\times10^{-14}\,  {\rm \ ph \ cm^{-2} \ s^{-1} \ GeV^{-1} }$, 
for $\alpha=0.5$, respectively.   


Sgr B2(M) is representative of a population  of molecular clouds in the Galactic center that display
 steady and
variable components of K$\alpha$ line emission as well as GeV and TeV $\gamma$-rays. Sunyaev \etal\, (1993) suggested
that the K$\alpha$ and nonthermal continuum emission from Galactic center molecular clouds is a result of an echo from
an X-ray flare from Sgr A*.  The variability analysis of some clouds showed superluminal motion and supported the
photo-ionization model of K$\alpha$ emission (Koyama \etal\, 1996; Terrier \etal\, 2010).  The high column density of
10$^{25}$ cm$^{-2}$ in Sgr B2(M) requires a special geometry in order to explain the X-ray echo model in Sgr B2(M).
The presence of a nonthermal radio source interacting with dense molecular gas in Sgr B2(M), suggests that a
nonthermal radio source interacting with molecular gas can also produce the 6.4 keV line emission with the assumption
that the metalicity in Sgr B2(M) is three times higher than the solar value.
Thus, it is not clear which mechanism dominates the K$\alpha$ line emission from Galactic center clouds. 
There are,
however, independent measurements favoring the cosmic ray picture, as described below. 
 
One is the critical role that low energy cosmic ray
electrons play in explaining the enhanced H$^+_3$ absorption implying high cosmic ray ionization rate (Oka \etal\,
2005). 
Second, the high energy $\gamma$-ray emission provides a strong 
constraint on the interaction of high energy cosmic
ray particles with Galactic center molecular clouds. 
The Galactic center is now recognized to have excess $\gamma$-ray emission  at energies between 20 MeV to TeV 
implying a  high cosmic ray density  (Aharonian \etal\,  2006; Abdo \etal\, 2009). 
Third, cosmic rays can naturally heat the gas and explain the elevated gas
temperature relative to the dust 
temperature in Galactic center molecular clouds (Yusef-Zadeh \etal\,  2007). 
Fourth, there is strong diffuse
nonthermal radio emission from the Galactic center as well as a large number of magnetized  radio 
filaments  (Nord \etal\, 2004; Yusef-Zadeh \etal\, 
2007a). In
particular,  74 MHz radio emission shows a spatial correlation with Galactic center molecular clouds
suggesting that cosmic ray electrons are diffusing through the  molecular gas (Yusef-Zadeh \etal\, 2013). Fifth,
the enhanced  abundance of SiO and methanol as well as the unusual abundance ratios found throughout Galactic center 
 molecular clouds
suggests that cosmic rays  drive  the chemistry of the gas (Yusef-Zadeh \etal\, 2013). Lastly, the variable radio emission observed on a $\sim$20-year time scale in Sgr B2(M) (De
Pree \etal\, 2014) may be due to the variable nonthermal radio source which is embedded within Sgr B2(M). 
Future monitoring of low frequency radio and
high energy emission will determine the contribution of cosmic rays in explaining the observed features in this
complex region of the Galaxy.

In summary, the  detection of OH(1720 MHz) maser emission adjacent to a nonthermal source suggests that 
a source of particle acceleration, most likely 
a SNR 
is physically interacting  with  the dense gas in  Sgr B2(M). 
These measurements have important implications for  the nature of 6.4 keV and $\gamma$-ray emission in the Galactic center, 
supporting a scenario in which cosmic rays are responsible for 
the production of high energy radiation  in Sgr B2(M).


Acknowledgments: This work is partially supported by the grant 
AST-1517246 from the NSF. The National Radio Astronomy Observatory is a 
facility of the National Science Foundation operated under cooperative 
agreement by Associated Universities, Inc. We thank the staff of the 
GMRT that made these observations possible. GMRT is run by the National 
Center for Radio Astrophysics of the Tata Institute of Fundamental 
Research.


\begin{figure}[p]
\centering
\includegraphics[scale=0.5,angle=0]{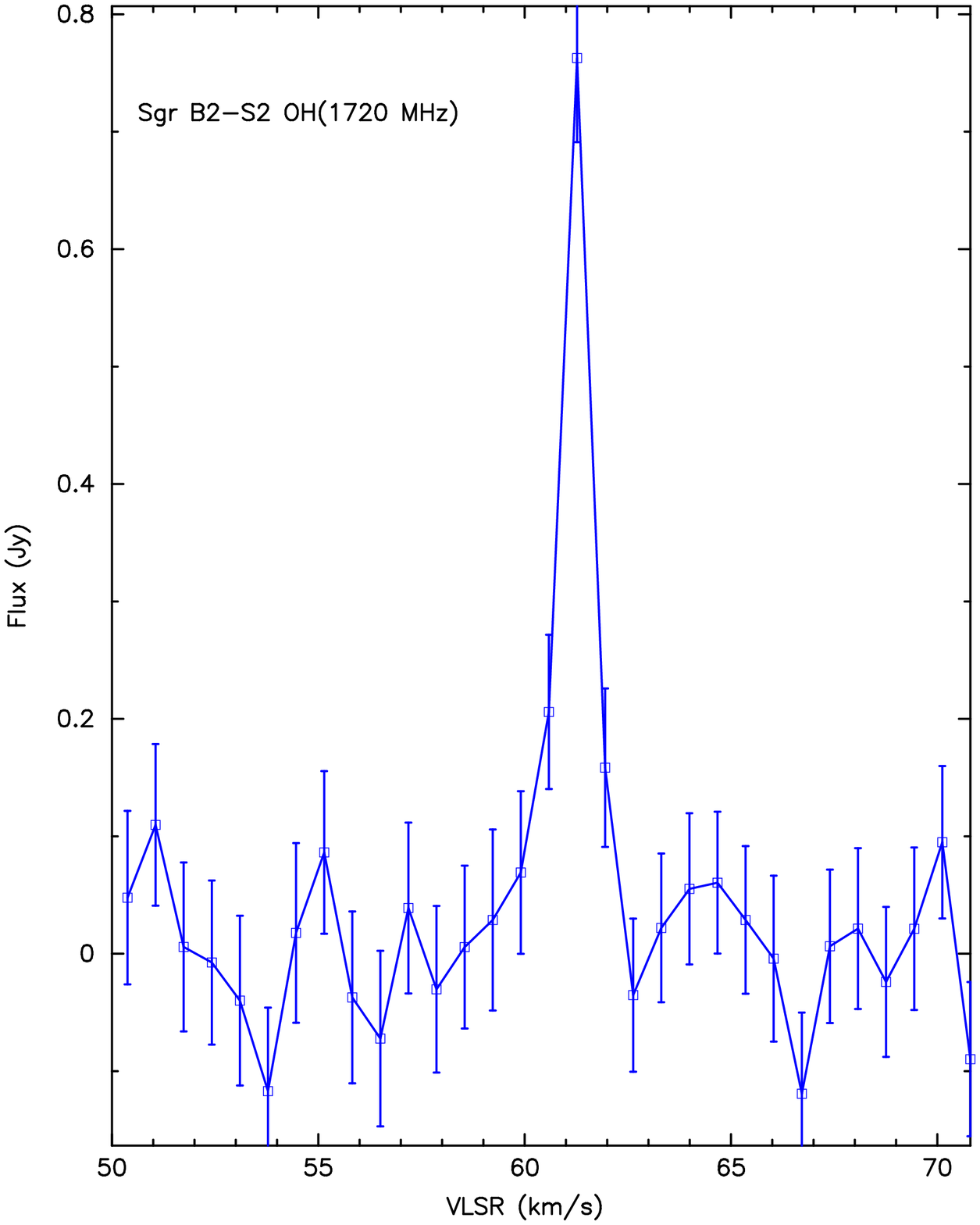}
\caption{An OH(1720 MHz)  spectrum toward G0.6647-0.0358 in Sgr B2(M). 
}
\end{figure}

\begin{figure}[p]
\centering
\includegraphics[scale=0.7,angle=0]{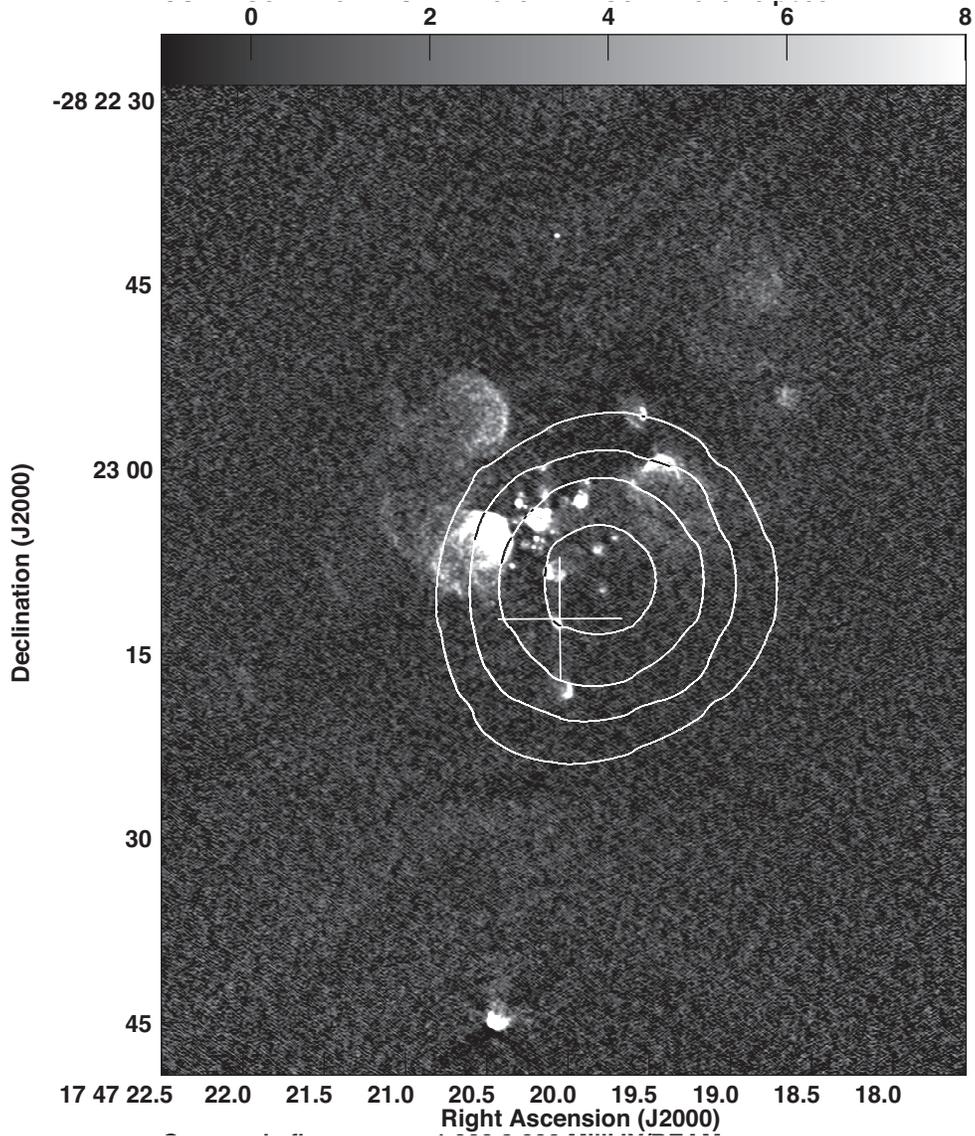}
\caption{
{\it (a)}
Contours of 150 MHz emission at 40, 60, 80, 100 and 120 mJy beam$^{-1}$ 
with a resolution of 25$''\times25''$ are superimposed on a 23 GHz  image of Sgr B2(M). 
with a resolution of  $0.36''\times0.18''$ (PA=2$^0$) (De Pree \etal\, 2005). 
The cross indicates the location of the  OH(1720 MHz) maser. 
{\it (b)} Similar to (a) except that contours of 327 MHz emission 
with 20, 40 and 60 mJy beam$^{-1}$ are superimposed 
on a grayscale image at 327 MHz 
with a resolution of 12.6$''\times6.8''$ (Nord \etal\, 2004). 
}
\end{figure}

\begin{figure}[tbh]
\centering
\includegraphics[scale=0.7,angle=0]{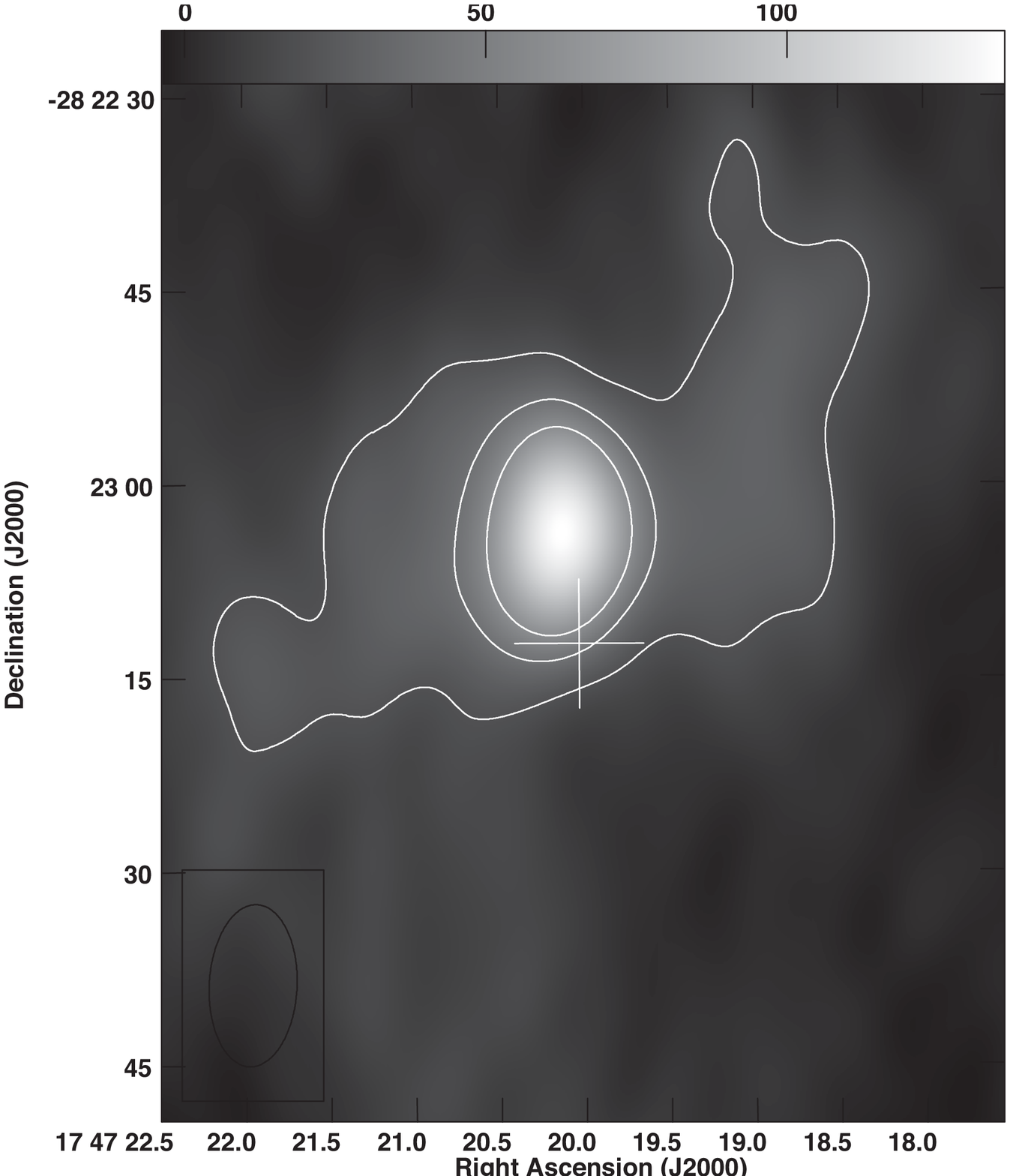}
\end{figure}

\begin{figure}
\centering
\includegraphics[scale=0.8,angle=0]{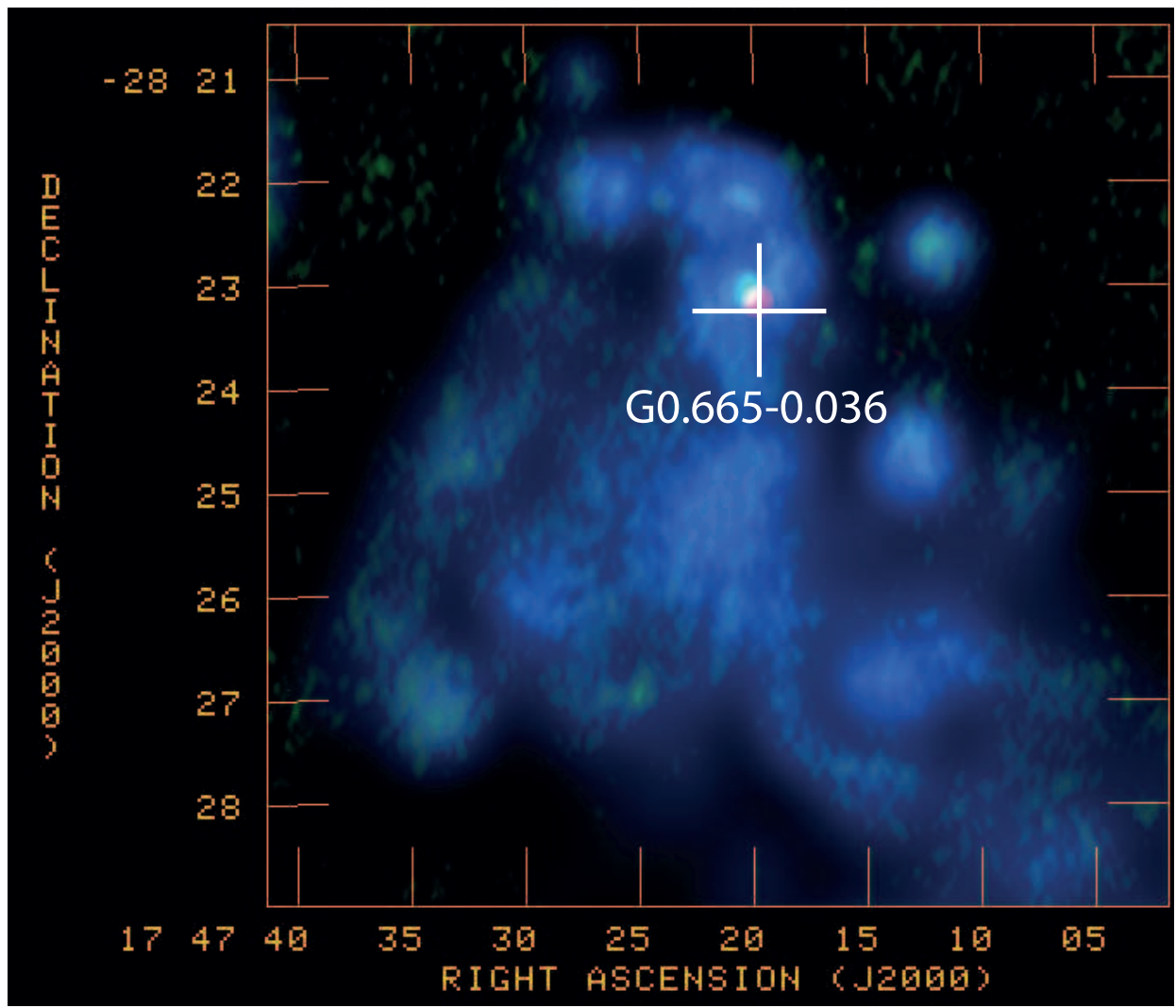}
\caption{
{\it (a)} 
A three color image of the 
Sgr B2 complex at 150 MHz, 327 MHz and 1.4 GHz in red, green and blue, 
respectively.The bright source coincides with Sgr B2(M) at the position 
where OH(1720 MHz) masers are detected. 
{\it (b) } 
Similar to (a) except the 
150, 255 and 327 MHz peaks (red, green, blue) are displayed with a resolution of 25$''$.  
The white spot coincides with the position of                
the nonthermal source.
}
\end{figure}

\begin{figure}[tbh]
\centering
\includegraphics[scale=0.8, angle=0]{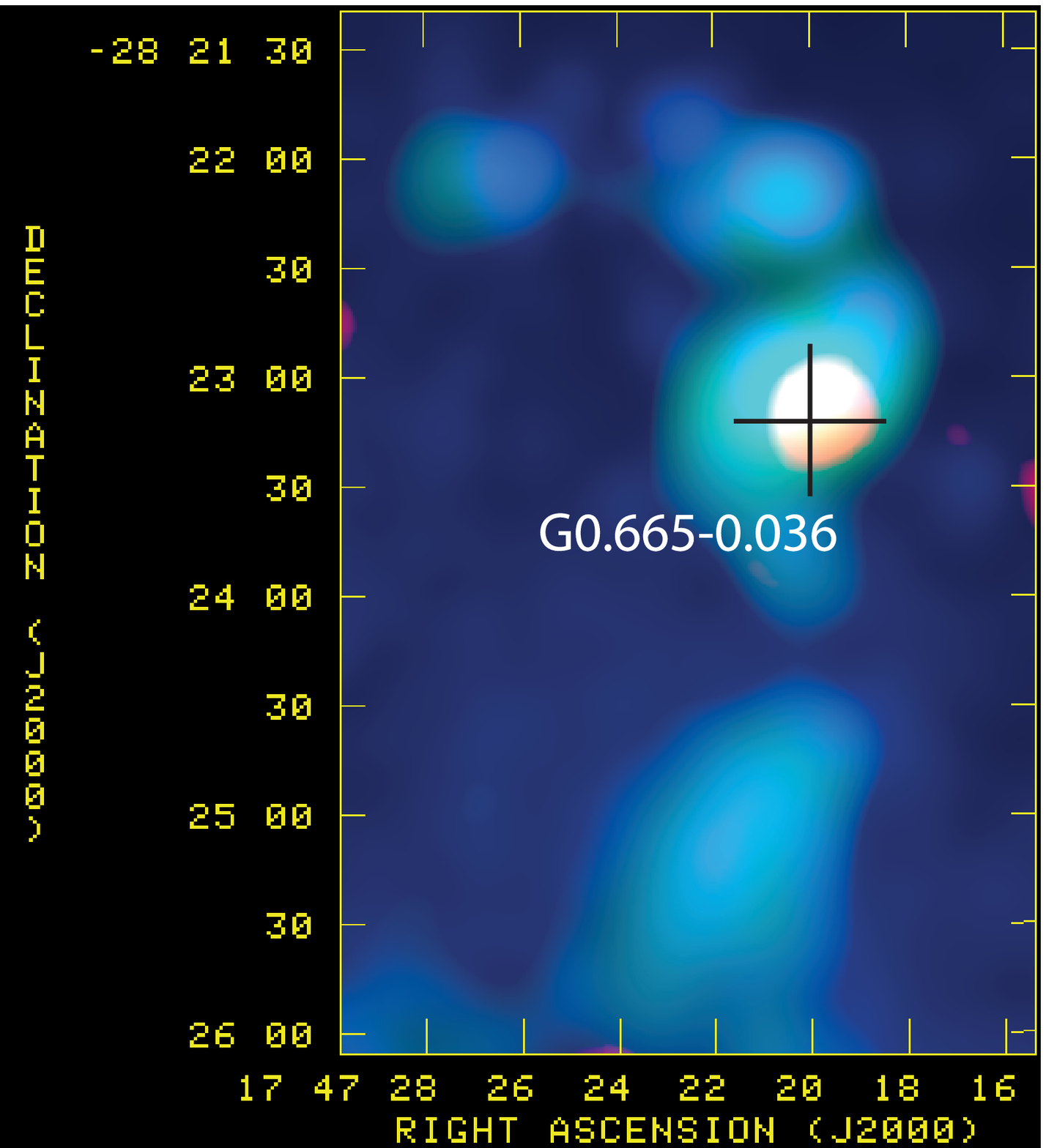}
\end{figure}

\begin{figure}[p]
\centering
\includegraphics[scale=1.5,angle=0]{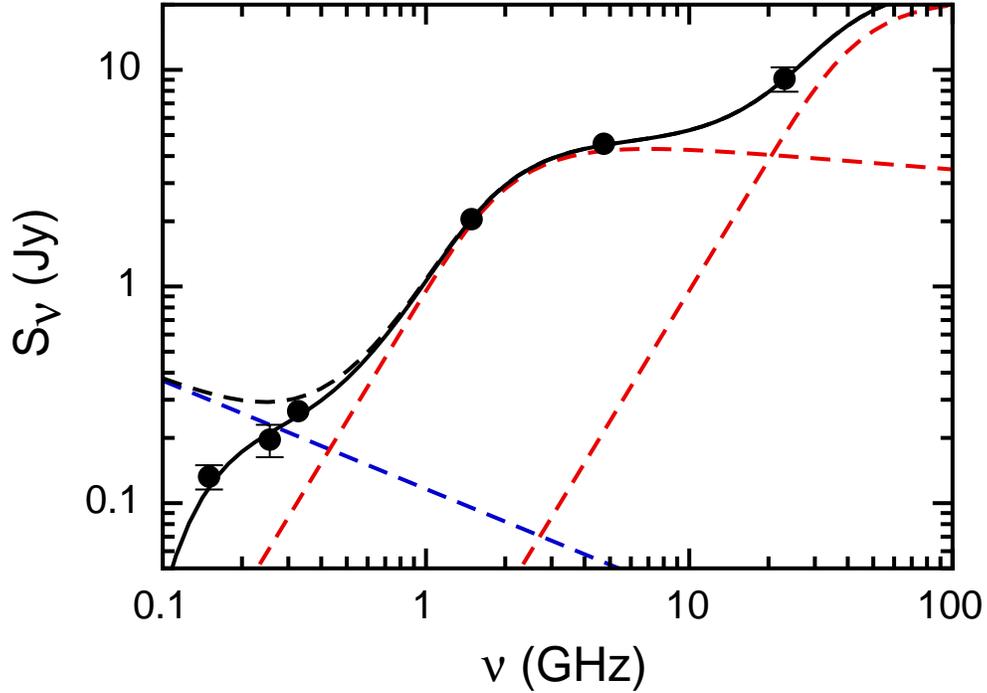}
\caption{
Observed (points) and modeled (solid curve) spectrum from Sgr B2. 
The observed peak flux density is estimated from 
background subtracted Gaussian fit to multi-wavelength images convolved to  
25$''$ resolution. 
Flux density 
errors for the three data points at intermediate frequencies are smaller
than the point size and are suppressed.  Dashed curves indicate the
inferred emission from the SNR  (blue), from compact and
UC HII regions within the beam (red), and the sum of all
three components (black).  The solid curve is obtained from the black
dashed curve by correcting for absorption by a thermal foreground with
$\tau=1$ at 150\,MHz (see text). 
}
\end{figure}

\end{document}